\documentclass[11pt]{article}
\usepackage{moriond,epsfig}

\def\be{\begin{equation}}
\def\ee{\end{equation}}
\def\bea{\begin{eqnarray}}
\def\eea{\end{eqnarray}}

\def\Mpc{\, h^{-1} \, {\rm Mpc}}

\def\Mpc{\ifmmode {\, h^{-1} \, {\rm Mpc}}
\else {$h^{-1}\,$ Mpc}\fi}

\def\s8{{\sigma_8}}

\def\ltsima{$\; \buildrel < \over \sim \;$}
\def\simlt{\lower.5ex\hbox{\ltsima}} 
\def\gtsima{$\; \buildrel > \over \sim \;$} 
\def\simgt{\lower.5ex\hbox{\gtsima}}

\def\omegam{{\Omega_{\rm m}}}
 
\def\omegab{{\Omega_{\rm b}}}
 
\def\omegal{{\Omega_\Lambda}}
 
\def\omegabh2{{\omegab h^2}} 

\begin{document}
\vspace*{4cm}
\title{Moriond Conference Summary: The Cosmological Model(s)}

\author{Ofer Lahav }

\address{Institute of Astronomy, University of Cambridge, Madingley Road\\
Cambridge CB3 0HA, England;  email:lahav@ast.cam.ac.uk}

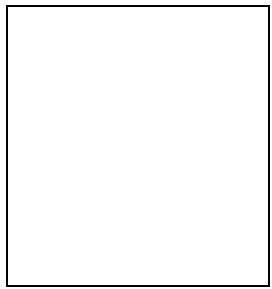
\maketitle\abstracts{The XXXVIIth Rencontres de Moriond on "The
Cosmological Model" is briefly summarized.  Almost none of the current
observations argues against the popular Cold Dark Matter + $\Lambda$
concordance model.  However, it remains to be tested how astrophysical
uncertainties involved in the interpretation of the different data sets
affect the derived cosmological parameters.
Independent tests are still required to establish if the Cold Dark Matter
and Dark Energy components are `real', or just `epicycles' that happen
to fit the current data sets well.}

\section{Introduction}

It is a challenging task in `data compression' to summarize
briefly a conference so rich in ideas and observational results, 
covered in over 100 oral presentations.
On the observational side, we were fortunate to hear at this meeting
for the first time the results from two CMB experiments: CBI and VSA,
which largely confirmed earlier results for the CMB acoustic peaks.  
We also heard updates on redshift and cluster surveys at different
wavelengths.  On the theoretical side, we learnt about the most advanced 
numerical simulations, and on ideas
which relate fundamental physics (e.g.`Brane World') to
cosmological models.  The exponential growth of data has changed the
character of the subject, in the sense that models can now be
assessed quantitatively in great detail. 

Below is a modest attempt to summarize the approaches of estimating
the `best fit' cosmological model.  It is interesting that the title
of meeting is ``The Cosmological Model'', perhaps implying the good
agreement within the community on the  concordance $\Lambda$-Cold Dark
Matter model.  As this model has been so successful and popular, 
it is timely to ask `what can go wrong', and if other models are still 
possible.


\section{Basic Assumptions and Paradigms}

As Jim Peebles emphasized in his Introduction talk, our study of the
Universe is  performed within the framework of  General Relativity,
and within that theory by assuming the 
Cosmological Principle, 
i.e. that on large scales 
the Universe is (roughly) homogeneous and isotropic.

It is worth noting that the Cosmological Principle and the resulting
FRW metric were formulated before observations could probe to
significant redshifts, when the `dark matter' problem was not
well-established and the Cosmic Microwave Background (CMB) was still
unknown.  However, this `guess' turned out to be successful.  The COBE
measurements of temperature fluctuations $\Delta T/T = 10^{-5} $ on
scales of $10^\circ$ give, via the Sachs Wolfe effect, rms density
fluctuations of ${{\delta \rho} \over {\rho}} \sim 10^{-4} $ on $1000
\Mpc$, i.e. the deviations from a smooth Universe are tiny.  However,
we note that on scales of $\sim 100 \Mpc$ (probed e.g. by the recent
redshift surveys 2dFGRS, SDSS) the rms fluctuations expected from
conventional models are non-negligible, ${{\delta \rho} \over {\rho}}
\sim 0.1$.  One also has to verify carefully that a sample properly
represents a typical patch of the FRW Universe in order to yield reliable
global cosmological parameters.
It is also common to assume the Inflationary paradigm, but other
Brane-inspired models have also gained popularity recently.

\section {Cosmic Probes}

The different cosmic probes discussed 
at the meeting determine different sets of cosmological parameters.
Below we give the approximated dependence on the density parameters
for matter (total) $\omegam$, 
baryons $\omegab$,  
and dark energy
$\omegal$, 
the Hubble constant
$H_0 \equiv 100 h$ km/sec/Mpc, 
and the (linear theory) 
normalization $\sigma_{8}$ of the mass fluctuations in $8 \Mpc$ spheres.
Other important parameters include the primordial spectral index $n$,
the age of the Universe $t_0$,  the optical depth to reionization $\tau$,
and the linear biasing parameter $b$ (which can be generalized to more
complicated biasing schemes). 
While in the 1960's
Sandage's goal was to `search for two numbers' it is now appreciated
that even the basic cosmological model requires a dozen or so
parameters, which are linked in non-trivial ways:

\begin{itemize}

\item Big Bang Nucelosynthesis: 
$\omegab h^2$

\item The CMB: 
$\omegal + \omegam, \;
\omegam h^2, \;\omegab h^2,\;   
\sigma_8 \; e^{-\tau},  \;n, \;t_0$

\item SN Ia:  
$3 \omegal - 4 \omegam$

\item Redshift surveys:  
$\omegam^{0.6}/b, \; b \sigma_8, \;\omegam h$

\item Peculiar velocities:
$\sigma_8 \omegam^{0.6}, \;\omegam h$

\item Cluster abundance:
 $\sigma_8 \omegam^{0.6}$

\item Weak lensing:
$\sigma_8 \omegam^{0.6}$

\item Baryon fraction in clusters:
$\omegab, \omegam, \; h$

\item Cepheids, Sunyaev-Zeldovich clusters, time-delay: $h$

\end{itemize}

An important point to note is that by using `orthogonal' constraints one can 
significantly improve the estimation of cosmological parameters.
The dependence on the factor of roughly $\omegam^{0.6}$ in different probes
(peculiar velocities, cluster abundance and cosmic shear) 
is a coincidence, but   
having a number of different probes which are sensitive 
to the same combination of parameters provides
an important cross check.

We also note that the choice of the model parameter space is somewhat arbitrary
(e.g. by fixing some parameters given priors).

Some of these measurements are easy to interpret.  For example the
physics of the CMB is linear, so the acoustic peaks at redshift $z
\sim 1000$ can probably be better  understood than the weather
conditions on the top of the Mont Blanc!  The interpretation of other
probes involves more complicated astrophysics, for example galaxy
biasing, the mass-temperature relation in the cluster abundance, the
evolution of SN Ia, and non-linear effects in weak gravitational 
lensing (shear).  
We also note that some of the methods (e.g. SN Ia and baryon fraction) are
free of assumptions on the nature of the dark matter,
while the interpretation of others (e.g. the CMB and redshift surveys)
requires a specific power spectrum of fluctuations (e.g. CDM) to be assumed.

\section{The Best-Fit Concordance Model: Ranking Our Belief}

Although the $\Lambda$-CDM model
with comparable amounts of Dark Matter and Dark Energy is rather esoteric,
it is remarkable that different measurements
converge to a `concordance model'.
However, some parameters are more robust than others,
and the list below is ordered (with a personal bias) 
from the more established 
to the less accurate parameters.
We note that the meaning of the error bars below is non-trivial, 
as in some cases they are the result of marginalization over other
free parameters.
In other cases the error bars reflect the diversity of  
results derived by different methods. 
Refined values and error bars can be found in contributions by others
in this volume and in numerous papers on astro-ph.

\begin{itemize}

\item The curvature of the Universe $\Omega_{\rm k} =
1 - \omegam - \omegal = 0 \pm 10\%$  
from the position of CMB acoustic peaks.

\item The baryon density $\omegab h^2 = 0.02 \pm 10 \% $ from BBN 
and the CMB.

\item The Hubble constant $h = 0.7 \pm 10 \% $ 
from the Cepheid-calibrated distances. 

\item  The spectral index $n  =  1 \pm 10\% $ from the CMB.

\item 
The age of the Universe $t_0 \approx 14 \pm 10\% $ Gyr from the CMB alone
(assuming $\Omega_{\rm k} = 0$).

\item The amplitude of fluctuations 
$\sigma_{8} \approx 0.8 \pm 30 \% $,
where the large error reflects the spread in results derived from 
cluster abundance, the CMB+2dFGRS and cosmic shear.

\item The mass density $\omegam  \approx 0.3 \pm 50 \%$,
where again the derived values vary considerably (e.g. from mass-to-light  
vs. velocity fields).

\item The neutrino mass density 
$ 0.001 < \Omega_{\nu} < 0.04 $, 
where the lower limit is from the recent atmospheric 
and solar neutrino oscillations and the upper limit 
is from large scale structure (e.g. 2dFGRS).

\item The cosmic equation of state $w = P/\rho$: 
the current data are consistent 
with $w=-1$ (i.e. a non-zero  Cosmological Constant), but different forms 
of Quintessence $w(z)$ are still possible.

\end{itemize}

There is still room for tensor modes, non-adiabatic and
non-Gaussian components in the CMB and  for better estimation of the
reionization parameter $\tau \sim 0.05$.  
There  is also the possibility of
time-dependent physical constants (e.g the fine-structure constant $\alpha$).
Conceptually, it seems we have to learn to live in a multi-component
complex Universe, which perhaps takes us away from 
an idealized model motivated by Occam's razor.

\section{Outlook}

There is general acceptance (perhaps too strongly) of the `concordance'
model with the following ingredients: 4\%  baryons, 26 \% Cold Dark
Matter (possibly with a small contribution of massive neutrinos) and
the remaining 70 \% in the form of Dark Energy (the Cosmological Constant
or `Quintessence').  
While phenomenologically the $\Lambda$-CDM
model has been successful in fitting a wide range
of cosmological data, there are some open questions:

\begin{itemize}

\item
Both components of the model, $\Lambda$ and CDM, 
have not been directly measured.
Are they `real' entities or just `epicycles'?
Historically  epicycles were actually quite useful 
in forcing observers to improve their measurements and theoreticians to think
about better models!

\item
`The Old Cosmological Constant problem': 
Why is $\Omega_\Lambda$ at present so small relative to what is expected
from Early Universe physics? 

\item 
`The New Cosmological Constant problem': 
Why is $\Omega_m \sim \Omega_\Lambda$ at the present-epoch?
Do we need to introduce  a new physics 
or to invoke the Anthropic Principle to explain it?

\item
There are still open problems in 
$\Lambda$-CDM on the small scales
e.g. galaxy profiles and satellites.

\item 
The age of the Universe is uncomfortably close to some estimates for the 
age of the Globular Clusters, if their epoch of 
formation was late. 

\item  
Could other (yet unknown) models fit the data equally well?

\item
Where does the field go from here?
Should the activity focus on refinement of 
the cosmological parameters within $\Lambda$-CDM, 
or on introducing entirely new paradigms?

\end{itemize} 

These issues will no doubt 
be revisited soon with larger and more accurate data sets.
We will soon be able to map
the fluctuations with scale and epoch, and to analyze jointly
redshift surveys 
(e.g. 2dF, SDSS, DEEP2) and
CMB (e.g. MAP, Planck) data. 
These high quality data sets 
will allow us to study a wider range of models and parameters.



\section*{Acknowledgments}
I thank the organisers for arranging this stimulating meeting,
and Marie Treyer for encouraging me to summarise it.
I am grateful to 
Sarah Bridle, Jerry Ostriker and other collaborators  in the 
Leverhulme Quantitative Cosmology group in Cambridge for  
helpful discussions.

\end{document}